\documentclass[aip,reprint,graphicx]{revtex4-1}
\usepackage{graphicx}
\usepackage{bm}
\usepackage{amssymb,amsmath}
\usepackage{braket}
\usepackage{array}
\usepackage{mathrsfs}
\usepackage{color}
\begin{document}

\title{Landau quantization in tilted Weyl semimetals with broken symmetry} 

\author{L. Zhang}
\affiliation{School of Physics, Georgia Institute of Technology, Atlanta, Georgia 30332, USA}
\author{Y. Jiang}
\email{yjiang@magnet.fsu.edu}
\affiliation{National High Magnetic Field Laboratory, Tallahassee, Florida 32310, USA}
\author{D. Smirnov}
\affiliation{National High Magnetic Field Laboratory, Tallahassee, Florida 32310, USA}
\author{Z. Jiang}
\email{zhigang.jiang@physics.gatech.edu}
\affiliation{School of Physics, Georgia Institute of Technology, Atlanta, Georgia 30332, USA}

\date{\today}

\begin{abstract}
Broken symmetry and tilting effects are ubiquitous in Weyl semimetals (WSMs). Therefore, it is crucial to understand their impacts on the materials' electronic and optical properties. Here, using a realistic four-band model for WSMs that incorporates both the symmetry breaking and tilting effects, we study its Landau quantization and the associated magneto-absorption spectrum. We show that the Landau bands in tilted WSMs can be obtained by considering a non-tilt Hamiltonian through Lorentz boost. However, broken symmetry effects can generate an additional term in the Hamiltonian, which equivalently leads to band reconstruction. Our work provides a more realistic view of the magnetic field response of WSMs that shall be taken into account in relevant future device applications.
\end{abstract}


\maketitle 

\section{Introduction}
The rise of Weyl semimetals (WSMs) \cite{WSM_T_1,WSM_T_2,WSM_E_1,WSM_E_2} is a boon to quantum materials research, as it not only hosts low-energy excitations that mimic the Weyl fermions in quantum field theory but also features topologically protected states at room temperature for novel practical applications.\cite{WSM_review_1,WSM_review_2,WSM_review_3,WSM_review_4} The intriguing properties of WSMs include the Fermi arcs,\cite{WSM_E_1,WSM_E_2,WSM_Arc_1,WSM_Arc_2,WSM_Arc_3,WSM_Arc_4} the ultrahigh mobility,\cite{WSM_M} the chiral anomaly effect,\cite{WSM_CA_1,WSM_CA_2,WSM_CA_3,WSM_CA_4,WSM_CA_5} the mixed axial-gravitational anomaly,\cite{WSM_GA} and the enhanced photovoltaic effect.\cite{WSM_PV_1,WSM_PV_2,WSM_PV_3,WSM_PV_4} These properties stem from the unique electronic structure of WSMs, consisting of linearly dispersing conduction band (CB) and valence band (VB) touching at two distinct Weyl points (WPs) in momentum space. Such a two-Dirac-cone-like (Weyl-cone) band structure has been the basis for revealing the WSM physics.

However, the above simple model does not capture the full aspects of Weyl physics. When the separation between WPs is small, they no longer can be treated as isolated WPs, and the interaction between Weyl cones has to be taken into account.\cite{Koshino_1} Consequently, a hybridization gap occurs when the two Weyl cones cross each other, leading to a reduction of symmetry. Such broken symmetry effects are responsible for the band gap opening at high magnetic fields and Weyl annihilation,\cite{WSM_transport_Jia} as well as a new selection rule for inter-Landau-band optical transitions.\cite{WSM_Yuxuan} These unique electronic and optical properties distinguish WSMs from Dirac semimetals. Another important effect is the tilting of the Weyl cones.\cite{WSM_tilt,WSM_titl_Trivedi} Although a small tilt does not qualitatively change the transport behaviors, it reconstructs the Landau bands and gives rise to new optical transitions beyond the usual electric dipole ones.\cite{Yang,Udagawa,Goerbig_1,Koshino_2} When the tilting is sufficiently large, the Fermi level ($E_F$) can intersect both the CB and VB simultaneously. Such an electron-hole coexisting system is thought to be a solid-state realization of breaking Lorentz invariance, as it cannot be adiabatically connected into a Lorentz conserved Hamiltonian.\cite{WSM_tilt,WSM_tilt_ZSY,WSM_tilt_Hasan}

Even though both the broken symmetry effects and the tilting effects are well established, few analyses have been done on their interplays. In this work, we employ a four-band effective Hamiltonian model that incorporates the WP coupling and band asymmetry of realistic WSMs to study the tilting effects in Landau quantization and magneto-absorption. We specifically focus on the titled Weyl bands in the archetypal WSM family of nonmagnetic transition-metal monopnictides (TX: T = Ta, Nb; X = As, P), as their electronic structures have been actively probed by magneto-infrared spectroscopy\cite{WSM_Yuxuan,WSM_IR_Xiu,WSM_IR_Orlita,WSM_IR_Jenkins} and their band parameters can be accurately determined via fitting the four-band model to the {\em ab initio} results. Our analysis shows that (i) the titled Weyl band can always be modeled by an equivalent, non-tilt Hamiltonian with a perfect match in Landau band energy; (ii) broken symmetry effects can lead to an additional term in the non-tilt Hamiltonian that breaks Lorentz invariance even with a small tilting angle; and (iii) the combined tilting and broken symmetry effects give rise to rich spectral features in magneto-absorption that could be mostly captured by a non-tilt model.

\section{Theoretical model}
We start with a $4\times 4$ Hamiltonian\cite{Koshino_1} that provides an excellent description of the low-energy electronic structure of coupled WPs in realistic WSMs\cite{WSM_review_4,WSM_Yuxuan}
\begin{equation}
H_0=T(\mathbf{k})+\tau_x\hbar(\boldsymbol{\sigma}\cdot\mathbf{vk})+ m\tau_z+b\sigma_x.
\label{Ham_1}
\end{equation}
Here, we assume the WPs are along the $x$ direction, $\mathbf{v}=(v_x,v_y,v_z)$ is an anisotropic band velocities, $\mathbf{k}$ ($\hbar\mathbf{k}$) is the wave (momentum) vector, and $\mathbf{vk}=(v_xk_x,v_yk_y,v_zk_z)$. In addition, $m$ is the mass parameter, $b$ describes the intrinsic Zeeman effect in the material, $\hbar$ is the reduced Planck constant, and $\sigma$ and $\tau$ denote the Pauli matrices for spins and orbitals, respectively. The tilting of the Weyl cones is quantified using $T(\mathbf{k})=\hbar(t_xv_xk_x\tau_x+t_yv_yk_y+t_zv_zk_z)$, where $\mathbf{t}=(t_x,t_y,t_z)$ is the tilt parameter vector. This form of $T(\mathbf{k})$ preserves the mirror symmetry of the Weyl band.\cite{WSM_Garnett} For nonmagnetic transition-metal monopnictides WSMs, {\em ab initio} calculations show that the largest tilt is along the $z$ direction with $t_z\approx 0.5$.\cite{abinitio_2} In the following Section, for simplicity, we will set the tilting along the $z$ direction only.\cite{note1}

\subsection{Lorentz boost and Landau bands}
\begin{figure}[t!]
\includegraphics[width=8.5cm]{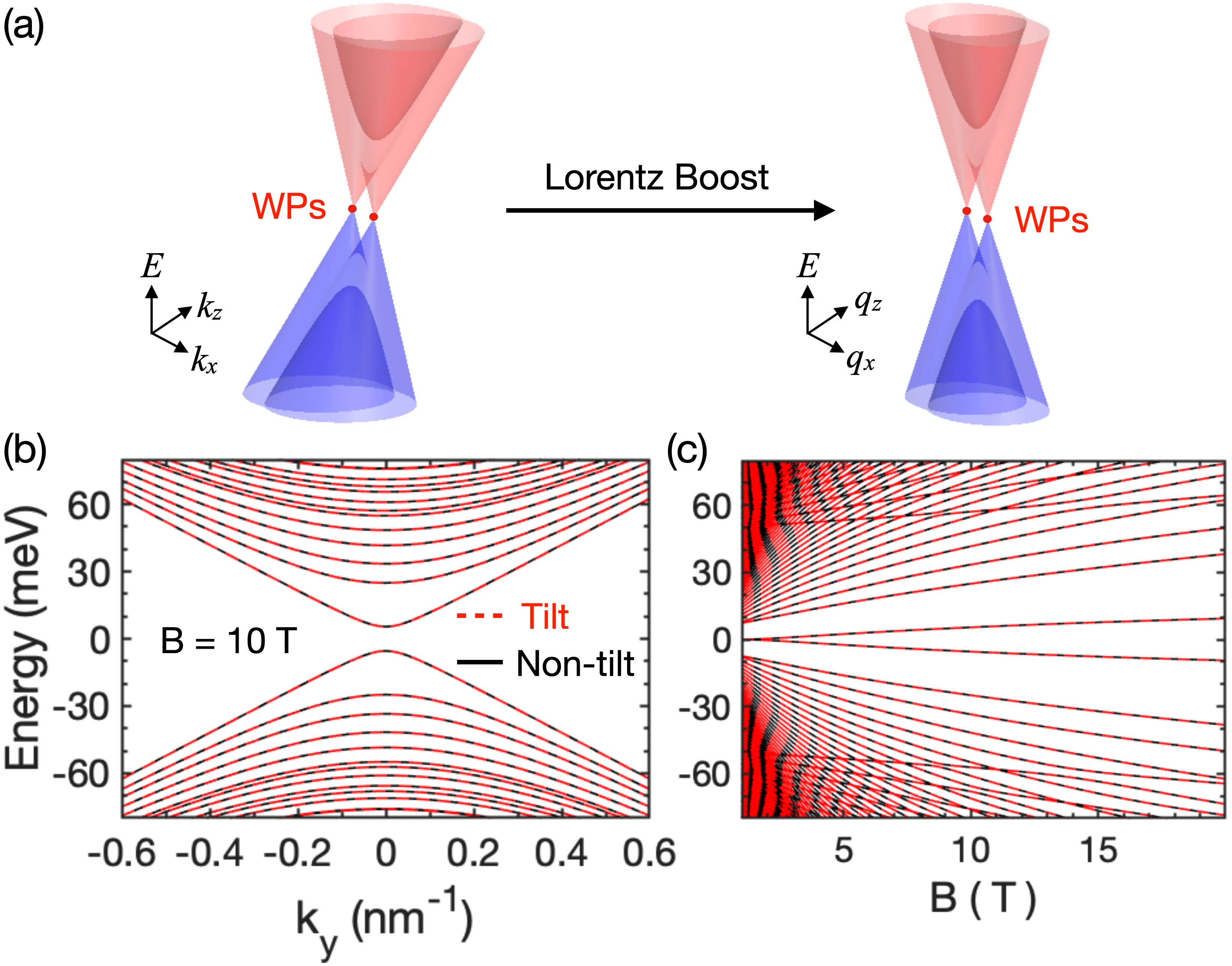}
\caption{(a) Schematic demonstration of Lorentz boost from tilted Weyl cones to non-tilt Weyl cones. (b) Landau band dispersion along the $k_y$ direction at $B=10$ T, applied along the $y$ direction. (c) Magnetic field dispersion of the Landau bands at $k_y=0$. A perfect match in Landau band energy between the tilt and non-tilt models is evidenced in both (b) and (c). All calculations are performed using the band parameters of WSM NbP enlisted in Table I.}
\end{figure}
It has been shown previously that the Landau bands of tilted WP can be obtained by finding an equivalent, non-tilt Hamiltonian through a Lorentz boost.\cite{Yang,Udagawa,Goerbig_1} This property reflects the relativistic nature of WSMs, enabling straightforward interpretations of experimental data. Here, we apply the Lorentz boost method to the more realistic Hamiltonian $H_0$ to eliminate the tilt along the $z$-direction. The Lorentz boost is done via the following transformation
\begin{align*}
    H_0'=\ &e^{\tau_x\sigma_z\theta/2}H_0 e^{\tau_x\sigma_z\theta/2}\\
    =\ &\tau_x\hbar(v_x \sigma_x k_x+v_y \sigma_y k_y)+\tau_x\hbar v_z \sigma_z k_ze^{\tau_x\sigma_z \theta}\\
    &+m\tau_z+b\sigma_x+Te^{\tau_x\sigma_z \theta}.
\end{align*}
By setting
\begin{align*}
    \beta_z=\tanh(\theta)=-t_z,\ 
    \gamma_z=\cosh(\theta)=\frac{1}{\sqrt{1-\beta_z^2}},
\end{align*}
we obtain
\begin{align}
\label{nontilt}
\begin{split}
    H_0'=\ &\tau_x\hbar(v_x \sigma_x q_x+v_y \sigma_y q_y)+\tau_x\hbar v_z \sigma_z q_z\\
    &+m\tau_z+b\sigma_x,
    \end{split}
\end{align}
where 
\begin{align*}
q_x=k_x,\quad q_y=k_y,\quad q_z=\frac{k_z}{\gamma_z}.    
\end{align*}
In this new coordinate system ($q_x, q_y,q_z$), the $T(\mathbf{k})$ term vanishes, and the Hamiltonian $H_0'$ leads to a non-tilt band structure. The concept of Lorentz boost is schematically shown in Fig. 1(a) for the case of $\beta_z=-t_z=-0.5$.

We note that since the Lorentz transformation is not unitary, $H_0$ and $H_0'$ represent two different energy bands at zero magnetic field. But, their Landau bands are equivalent with proper scaling. To see this, we solve the following two eigenvalue equations\cite{Goerbig_1}
\begin{align}
\label{eigen}
    (H_0-E)\ket{\Phi}=0,\\
    (H_0'-Ee^{\tau_x\sigma_z\theta})\ket{\Tilde{\Phi}}=0,
\end{align}
where $E$ is the energy, $\ket{\Tilde{\Phi}}=\mathcal{N}\exp(-\tau_x\sigma_z\theta/2)\ket{\Phi}$, and $\mathcal{N}$ is a normalization constant. Here, we set the magnetic field ($B$) along the $y$ direction, perpendicular to the WP separation direction ($\mathbf{k}_{\text{WP}}$) as an example, but the analyses and results are similar with magnetic field along the other directions. In this setting, we can again redefine a coordinate system $(q_x',q_y',q_z')$, where $q_x'=q_x$, $q_y'=q_y$, and $q'_z=q_z-\beta_z\gamma_zE/\hbar v_z$, and rewrite the momentum operators in $H_0$ and $H_0'$ with ladder operators. Using Landau gauge $\mathbf{A}=(0,0,-Bx)$, the ladder operators read $a=(\Pi_x+i\Pi_z)/\sqrt{2e\hbar B}$ and $a^\dagger=(\Pi_x-i\Pi_z)/\sqrt{2e\hbar B}$, where $e$ is the elementary charge, and the canonical momentum $\mathbf{\Pi}=\hbar \mathbf{k}+e\mathbf{A}$ in Eq. (3) and $\mathbf{\Pi}=\hbar \mathbf{q'}+e\mathbf{A}$ in Eq. (4), respectively. In this way, we find the Landau bands of the two Hamiltonians having an identical form if the following replacements are made
\begin{align}
\label{renorm}
\begin{split}
    m'&=\sqrt{1-\beta_z^2}\,m,\ b'=\sqrt{1-\beta_z^2}\,b,\\
    v'_{x,z}&=(1-\beta_z^2)^{3/4}v_{x,z},\ v'_y=\sqrt{1-\beta_z^2}\,v_y.
\end{split}
\end{align}
From now on, we will refer the tilt and non-tilt models to Eqs. (3) and (4) utilizing the Hamiltonian before and after Lorentz boost. 

We then numerically calculate and compare the Landau bands between the two models by expanding their wavefunctions in the basis of Hermite polynomials with a cutoff of 100 terms. We use a realistic set of band parameters of WSM NbP, $m=24$ meV, $b=34$ meV, $v_x=3.8\times 10^5$ m/s, $v_y=1.8\times 10^5$ m/s, $v_z=2.7\times 10^5$ m/s, and $\beta_z=-0.5$ (Table I). Figure 1(b) shows the Landau band dispersion along the $k_y$ direction at $B=10$ T, and Fig. 1(c) shows the magnetic field dispersion at $k_y=0$. A perfect match in Landau band energy between the tilt and non-tilt models is evidenced.

\subsection{Landau bands mismatch due to symmetry breaking}
It is important to note that the Lorentz boosted non-tilt Weyl band and the resulting perfect match in Landau band energy are not always guaranteed due to symmetry breaking.\cite{book_Grushin} In this Section, we will analyze the broken symmetry effects on Landau bands mismatch.

\subsubsection{Electron-hole asymmetry}
The above four-band model \eqref{Ham_1} preserves the electron-hole (e-h) symmetry except for the tilt term. However, in reality, asymmetric band edges occur in the CBs and VBs due to the e-h asymmetric intrinsic Zeeman effect, as evidenced in both angle-resolved photoemission spectroscopy (ARPES) measurements \cite{ARPES_eh_1,ARPES_eh_2} and band structure calculations.\cite{first_1,abinitio_1,abinitio_2} Therefore, we modify $H_0$ as
\begin{equation}
H_{eh}=H_0+\delta b \tau_z\sigma_x,
\label{Ham_eh}
\end{equation}
where $\delta b$ describes the level of e-h asymmetry (see Appendix I for details). Now the model not only provides a decent description around the WPs, but also captures the correct hybridization gap and dispersion for the upper CB and lower VB.

\begin{figure}[t!]
\includegraphics[width=8.5cm]{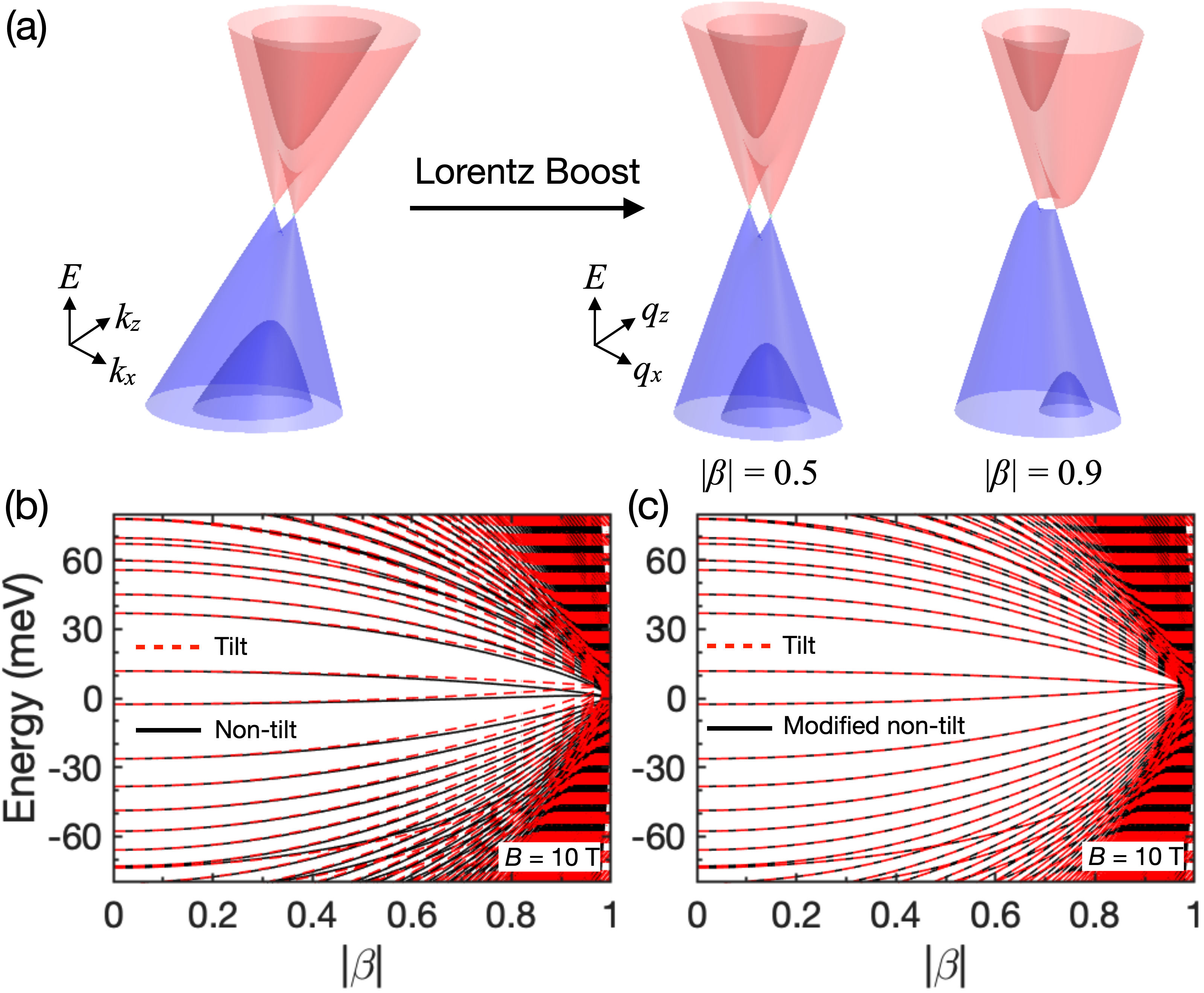}
\caption{(a) Schematic demonstration of Lorentz boost from e-h asymmetric tilted Weyl cones to non-tilt Weyl cones. Here, the symmetry in the boosted Weyl band is further reduced. (b) Landau bands matching at different tilting angles between the tilt model and the non-tilt model (using the renormalized band parameters \eqref{renorm}). (c) Same as (b) but comparing the tilt model with the modified non-tilt model (using $H_{eh}'$). The Landau bands are calculated at $k_y=0$ with the same setting as in Figs. 1(b) and 1(c) but including e-h asymmetry.}
\end{figure}

In carrying out Lorentz boost to $H_{eh}$, the inclusion of e-h asymmetry modifies the non-tilt Hamiltonian as  
\begin{align}
\label{modifyH}
\begin{split}
H_{eh}'=H_{0}'+\gamma_z\delta b (\tau_z\sigma_x+\beta_z\tau_y\sigma_y).
\end{split}
\end{align}
This result is in stark contrast to $H_0'$, where the Lorentz boost only eliminates the tilt term so that $H_0'$ exhibits a non-tilt form. Here, however, not only $\delta b \tau_z\sigma_x$ is scaled to $\gamma_z \delta b \tau_z\sigma_x$ after transformation, but more importantly, an additional term $\beta_z\gamma_z \delta b \tau_y\sigma_y$ emerges which is not present in its original Hamiltonian $H_{eh}$. Figure 2(a) visualizes how the boosted Weyl band is modified by the additional $\beta_z\gamma_z \delta b\tau_y\sigma_y$ term at zero magnetic field. As one can see, although the WPs still preserve, the upper CB and VB are both offset to the negative $k_z$ direction while the lower CB and VB shifted to the positive $k_z$ direction, breaking the symmetries within the $k_x$-$k_z$ plane. 

Consequently, the Landau bands of $H_{eh}$ with a tilt cannot be simply obtained by the method described before, that is, setting the tilt to zero and using the renormalized band parameters \eqref{renorm} together with $\delta b'=\sqrt{1-\beta_z^2}\,\delta b$. This additional term will inevitably modify the Landau bands regardless of the magnetic field directions, but the details depends on the magnitude of $\delta b$, field directions and also the tilting angles. Fig. 2(b) shows the matching between these two models in NbP with our current setting. Due to the prefactor $\beta_z$ in the additional $\tau_y\sigma_y$ term, the deviation in Landau band energy is not significant for small tilting angles. Noticeable differences only appear when $|\beta|>0.5$. Therefore, the above non-tilt model with renormalized band parameters could still work as a great approximation in e-h asymmetric WSMs with a small tilting angle, including the nonmagnetic transition-metal monopnictides interested herein. For large tilting angles, we can build a modified non-tilt model using $H_{eh}'$ to recover the contribution of the additional $\tau_y\sigma_y$ term and match the Landau bands of the tilt model. A perfect match in Landau band energy between the two models is shown in Fig. 2(c).

\subsubsection{Broken symmetry due to the intrinsic Zeeman effect}
The reason for the additional term in $H_{eh}'$ is because the symmetry breaking term $\delta b \tau_z \sigma_x$ in \eqref{Ham_eh} commutes with the transformation matrix $\tau_x\sigma_z$. Simply put, if the tilt is along a symmetry breaking direction, then the Lorentz boost transformation will generate an additional symmetry breaking term in the non-tilt Hamiltonian. Compared with the commonly used two-band model\cite{Goerbig_1,WSM_Carbotte} where the isolated WPs exhibit spherical symmetry, the intrinsic Zeeman effect term $b\sigma_x$ in our four-band model reduces it to the axial symmetry with respect to $\mathbf{k}_{\text{WP}}$. Therefore, in principle, we can consider the case of $\mathbf{t}\parallel\mathbf{k}_{\text{WP}}$ and Lorentz boost via transformation matrix $\sigma_x$ for $H_0$. Since both $m\tau_z$ and $b\sigma_x$ commute with $\sigma_x$, the transformed Hamiltonian now generates two additional terms after Lorentz boost
\begin{align*}
b\sigma_x\longrightarrow \gamma_x b(\sigma_x+\beta_x),\quad
m\tau_z \longrightarrow \gamma_x m(\tau_z+\beta_x\tau_z\sigma_x),
\end{align*}
where $\beta_x=\tanh(\theta)=-t_x$, and $\gamma_x=\cosh(\theta)$. However, in a magnetic field, the non-tilt model now fails to reproduce the correct Landau bands of the tilt model. This is because the off-diagonal $E$ term from $Ee^{\sigma_x\theta}$ can no longer be absorbed into one of the $k$. In reality, since the tilting along $\mathbf{k}_{\text{WP}}$ is not possible for all WSMs in the nonmagnetic transition-metal monopnictides family, we will not pursue further investigation here.

\subsection{Magneto-absorption}\label{absorption}
Having calculated the Landau bands and their wavefunctions, we can further compute the magneto-absorption coefficient $\alpha$ of WSMs using Fermi's golden rule\cite{YJ_PRB_InAsGaSb}
\begin{align*}
\begin{split}
    \alpha(E) \propto &\sum_{\nu} \sum_{n_1,n_2} \frac{B}{E} |\bra{n_1}v_{\nu}\ket{n_2}|^2 \\&\left(f(E_{n_1})-f(E_{n_2})\right)\delta(E- (E_{n_2}-E_{n_1})).
\end{split}
\end{align*}
Here, $v_\nu=\partial{H}/\partial{(\hbar k_\nu})$ is the velocity operator, $f$ is the Fermi-Dirac distribution, and $n_1$ and $n_2$ denote the initial and final Landau bands summing over all possible combinations. The index $\nu$ indicates the light polarization direction, which we set to be along the $x$ and $z$ directions ($\perp\mathbf{B}$). The magneto-absorption coefficient is proportional to the real part of the optical conductivity and can be measured directly using magneto-infrared spectroscopy. In the following discussion, we assume $E_F$ is pinned at the WPs, that is, at the charge neutrality, while leaving the discussion of finite doping effects in Appendix II. We also replace the delta function by a Lorentzian function with a  phenomenological broadening of 1 meV. Finally, since the experimentally observable optical transitions are dominated by the contributions at the band extrema, where the joint density of states diverges, we calculate the Landau band transitions at $k_y=0$. This is the only Van Hove singularity along the magnetic field direction in our setting ($\mathbf{B}\perp\mathbf{k}_{\text{WP}}$), as shown in Fig. 3(c).

\begin{figure}[t!]
\includegraphics[width=8.5cm]{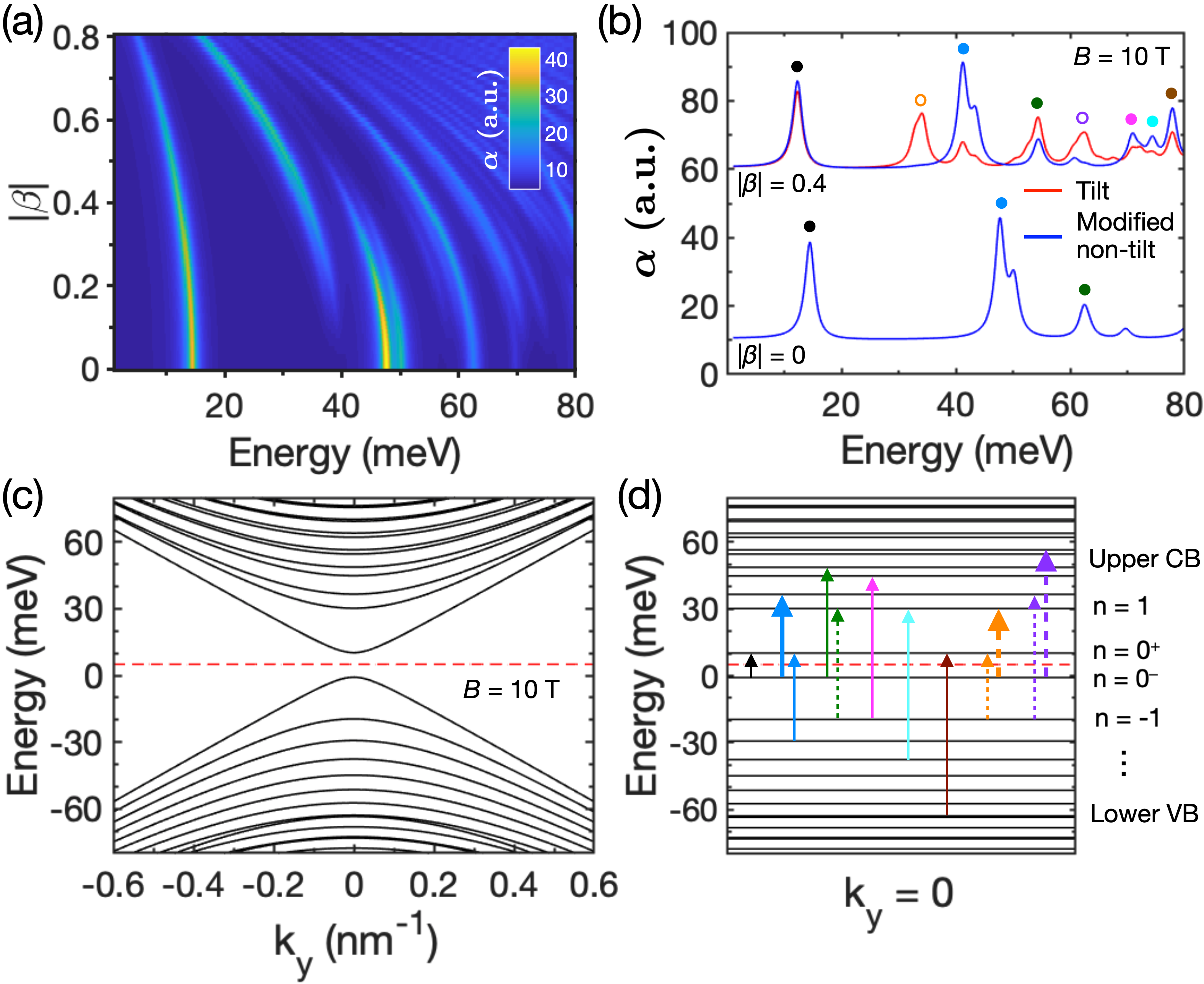}
\caption{(a) Calculated magneto-absorption spectra $\alpha(E)$ at $B=10$ T (using the tilt model) as a function of tilting angle. As expected, more transitions appear with increasing tilt. (b) Comparison of the tilt and modified non-tilt models at $|\beta|=0$ and $|\beta|=0.4$. The magneto-absorption spectra of the two models show a perfect match at zero tilt, but deviation occurs at a finite tilt. (c) Landau band dispersion along the $k_y$ direction at $B=10$ T. (d) Low-lying transitions at $k_y=0$ in (c), color-coded corresponding to the major absorption peaks in (b). The solid (dash) arrows in (d) as well as the solid (open) circles in (b) label the transitions at $|\beta|=0$ ($|\beta|=0.4$), respectively. For a split peak, the thick (thin) arrows mark the strong (weak) transitions. The red dash line in (c) and (d) indicates the Fermi level position.}
\end{figure}

Figure 3(a) shows the calculated magneto-absorption spectra $\alpha(E)$ at $B=10$ T and different tilting angles, using the tilt model. Figure 3(b) compares the detailed spectra for the tilt and modified non-tilt models at $|\beta|=0$ and $|\beta|=0.4$. Here, we neglect the non-tilt model because it produces similar spectra as the modified non-tilt model when the tilt is small ($|\beta|<0.5$). At zero tilt ($|\beta|=0$), the magneto-absorption spectra of all models overlap with each other and exhibit a series of interband transitions such as $L_{0^-(-2)}\rightarrow L_{2(0^+)}$ (light blue) and $L_{-1(-3)}\rightarrow L_{3(1)}$ (magenta and cyan), following a selection rule of $\Delta n\equiv|n_2|-|n_1|=\pm 2$.\cite{WSM_Yuxuan} The e-h asymmetry further splits the electron-like ($\Delta n>0$) and hole-like ($\Delta n<0$) transitions, as evidenced by the doublet structure of the $L_{0^-(-2)}\rightarrow L_{2(0^+)}$ (light blue) transition.\cite{note3} Additionally, several pronounced chiral transitions (that is, involving the $0^{\text{th}}$ LL) are observed, including the $L_{0^-}\rightarrow L_{0^+}$ (black) transition, the green transition from the upper VB to the upper CB, and the brown transition from the lower VB to the lower CB. All the observable transitions at $|\beta|=0$ are labeled by solid arrows in Fig. 3(d) and solid circles in Fig. 3(b). These transitions, along with the new selection rule, are the results of the WPs coupling.

As the tilt increases, the intensity of the aforementioned transitions decreases, and new transitions labeled by dash arrows in Fig. 3(d) and open circles in Fig. 3(b) emerge. All the transitions redshift due to the Fermi velocity renormalization. Since the high energy transitions are relatively weak and tend to merge to form a continuum background, we shall only focus on the experimentally discernible low-lying ones. We find that except for two new modes (orange and purple) and intensity mismatches, most transitions in the tilt model can be captured by the modified non-tilt model in Fig. 3(b), indicating its effectiveness as an approximation to the tilt model. Among those new transitions, chiral transitions are the most visible, particularly $L_{0^-(-1)}\rightarrow L_{1(0^+)}$ (orange), as it maintains strong intensity even at a large tilt. Therefore, it can be viewed as an experimentally accessible indicator for the tilted Weyl bands.

\section{Discussion}
The additional term generated from the Lorentz boost in the four-band model is indicative of Lorentz invariance violation. Indeed, the symmetry breaking terms $b\sigma_x$ and $\delta b \tau_z\sigma_x$ correspond to Lorentz-violating terms in high-energy theory.\cite{book_Grushin} These terms result in distinct optical responses and signify the differences between the Weyl and Dirac semimetals.

From an experimental point of view, magneto-infrared spectroscopy provides an effective way to identify the tilting effects in WSMs with several observables complimentary to the zero-field approach. \cite{tilt_carbotte} The first is the breaking of the zero-tilt selection rule, which leads to the observation of the otherwise forbidden transitions. When the tilt is relatively large ($|\beta|>0.6$), the higher-energy transitions tend to merge, causing an oscillatory behavior in the magneto-absorption spectra. The second is the renormalization of the band velocity and mass. In a magnetic field (taking $\mathbf{B}\parallel y$-axis), the Landau band energy near WPs scales with $\hbar \sqrt{v_x v_z} (1-\beta_z^2)^{3/4}/l_B$ (where $l_B=\sqrt{\hbar/eB}$ is the magnetic length),\cite{ZrTe5_YJ_PRL} whereas the mass scales with $m (1-\beta_z^2)^{1/2}$. Such $\beta_z$ dependence can be viewed as the consequence of tilting induced band renormalization described by Eq. \eqref{renorm}. We note that the magnetic field for these two energy scales being comparable gets larger with increasing tilt, making the magneto-infrared spectra of highly tilted WSMs less Dirac-like, that is, deviated from the characteristic $\sqrt{B}$ dependence.

\begin{table}[t!]
\caption{Band parameters of nonmagnetic transition-metal monopnictides WSMs extracted from the {\em ab initio} calculations of Ref. [\onlinecite{abinitio_2}]. Here, $x$ is the WP separation direction, and $z$ is the $c$ axis of the crystal. WP1 (WP2) refers to the coupled WPs lying on (off) the $k_z=0$ plane.}
\begin{center}
\begin{tabular}{|m{1.8cm}|c|c|c|c|c|c|c|c|}
\hline 
&\multicolumn{4}{c}{WP1} \vline &\multicolumn{4}{c}{WP2} \vline \\ \hline
\centering WSMs & TaAs & TaP & NbAs & NbP & TaAs & TaP & NbAs & NbP \\ \hline
\centering $m$ (meV) &76 &68 &21 &17 &83 &78 &25 &24 \\
\centering $b$ (meV) &104 &97 &35 &26 &131 &120 &39 &34\\
\centering $\delta b$ (meV) &-55 &-50 &-15 &-9 &-20 &-25 &-8 &-8\\
\centering $v_x$ ($10^5\ $m/s) &6.4 &6 &6.4 &6  &4.2 &4.4 &3.9 &3.8\\
\centering $v_y$ ($10^5\ $m/s) &2.3 &2.6 &2.4 &2.4 &3 &2.6 &1.5 &1.8\\
\centering $v_z$ ($10^5\ $m/s) &0.15 &0.3 &0.3 &0 &2.9 &2.7 &2.9 &2.7\\
\centering $t_x$ &0 &0 &0 &0 &0 &0 &0 &0\\
\centering $t_y$ &0.3 &0.35 &0.3 &0.4 &0.15 &0.15 &0.15 &0.1\\
\centering $t_z$ &0 &0 &0 &0 &0.52 &0.55 &0.49 &0.55\\
$\mathscr{B}_x$ (eV$\cdot$nm$^2$)&-0.1 &-0.1 &-0.1 & 0 &0 &0 &0 &0\\
$\mathscr{B}_y$ (eV$\cdot$nm$^2$)&-0.14 &-0.12 &-0.02 &-0.18 &-0.16 &-0.1 &-0.15 &-0.16\\
$\mathscr{B}_z$ (eV$\cdot$nm$^2$)&-0.01 &-0.01 &-0.05 &-0.04 &-0.1 &-0.2 &-0.18 &-0.2\\
\hline
\end{tabular}
\end{center}
\end{table}

Lastly, we want to make a few additional comments on the tilting effects in nonmagnetic transition-metal monopnictides WSMs. To facilitate the discussion, we first extract the band parameters via fitting the band structures of {\em ab initio} calculations \cite{abinitio_2} with the four-band model of \eqref{Ham_fit}. There are 12 pairs of coupled WPs in these materials.\cite{WSM_T_1,WSM_T_2,WSM_E_1,WSM_E_2} But, we can simply categorize them into two types, WP1 and WP2, for the fitting. The fitting procedure is described in Appendix I, and the resulting band parameters are listed in Table I. We note that for both WP1 and WP2, the tilt is not very large ($|\beta|<0.6$). The non-tilt model is thus applicable to both WPs. The broken symmetry effects (related to the $m$, $b$ and $\delta b$ terms) are in general larger in Ta-based WSMs than Nb-based ones, indicative of their correlation with the strength of spin-orbit coupling.\cite{WSM_Arc_3} Also, as mentioned above, since the tilt is in the directions perpendicular to the WP separation direction, only e-h asymmetry (related to the $\delta b$ term) would lead to a symmetry breaking term in the Hamiltonian. As $t_z=0$ for WP1, the broken symmetry effects only occur in WP2.

\section{Conclusion}
In summary, using a realistic four-band model, we show that a tilted Weyl band can be transformed into a non-tilt one through the use of Lorentz boost, which greatly facilitates the understanding of the magnetic field responses of WSMs. However, due to the broken symmetry effects such as the e-h asymmetry, the transformation can generate an additional term in the non-tilt Hamiltonian, which leads to a reconstructed Weyl band in magnetic fields. Using nonmagnetic transition-metal monopnictides WSMs as examples, we evaluate the impacts of the tilting and broken symmetry effects on the Landau quantization and magneto-absorption. We find that a simple non-tilt model with renormalized band parameters can effectively reproduce the Landau bands, while the resulting magneto-absorption spectra need to include transitions beyond the usual selection rule.

\begin{acknowledgments}
We thank Davide Grassano and Olivia Pulci for sharing their {\em ab initio} calculation results beyond Ref. [\onlinecite{abinitio_2}]. This work is primarily supported by the DOE through Grant No. DE-FG02-07ER46451. The work performed at the NHMFL is also supported by the NSF Cooperative Agreement No. DMR-1644779 and the State of Florida. Y.J. acknowledges support from the NHMFL Jack Crow Postdoctoral Fellowship, and Z.J. acknowledges the NHMFL Visiting Scientist Program.
\end{acknowledgments}

\section*{Appendix I: Extracting band parameters of nonmagnetic transition-metal monopnictides WSMs}
\begin{figure*}[t!]
\includegraphics[width=18cm]{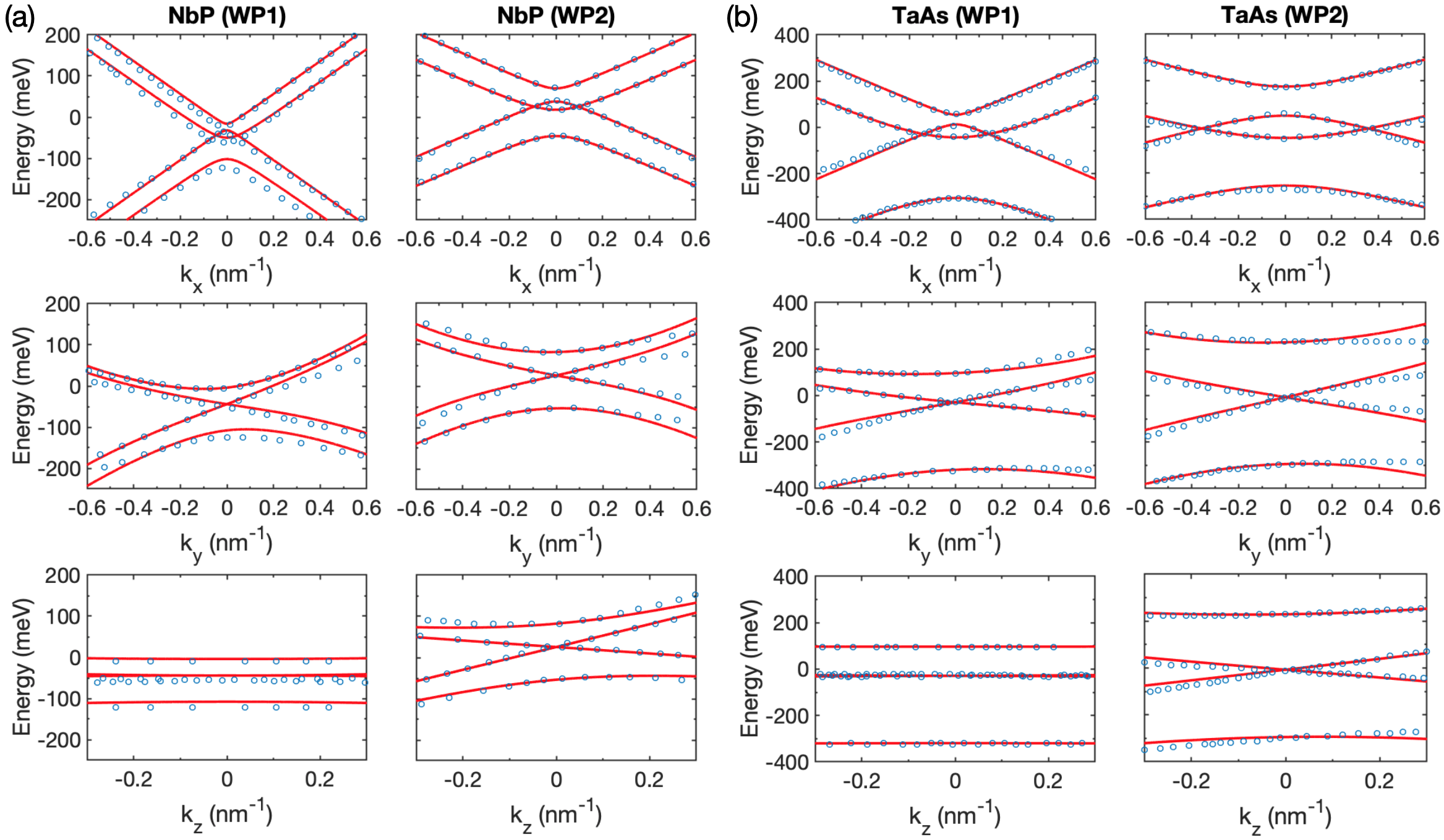}
\caption{Fitting results along different $k$ directions for both WP1 and WP2 in (a) NbP and (b) TaAs. The open circles (blue) are extracted from the {\em ab initio} calculations in Ref. [\onlinecite{abinitio_2}], while the solid lines (red) are the best fits using the four-band model \eqref{Ham_fit}. The fit along the $k_x$ direction is shown in the top row with $k_y=0$ and $k_z=0$, while that along the $k_y$ direction is in the middle row with $k_x=k_{\text{WP}}$ and $k_z=0$, and that along the $k_z$ direction in the bottom row with $k_x=k_{\text{WP}}$ and $k_y=0$. Here, $\pm k_{\text{WP}}$ represent the positions of the WPs in the $k_x$ direction.}
\end{figure*}
In this Appendix, we describe how we extract the band parameters of nonmagnetic transition-metal monopnictides WSMs via fitting the band structures of {\em ab initio} calculations \cite{abinitio_2} with the four-band model. For better presentation, we set the WP separation along the $z$ axis, which can be easily done by replacing $b\sigma_x$ with $b\sigma_z$ in Eq. \eqref{Ham_1}. Note that the band edges of the four bands at $k_z=0$ are solely determined by the two parameters $m$ and $b$ in Eq. \eqref{Ham_1}. To fully account for the asymmetric band edges observed in both ARPES measurements \cite{ARPES_eh_1,ARPES_eh_2} and band structure calculations,\cite{first_1,abinitio_1,abinitio_2} one can introduce an e-h asymmetry for both the mass $m$ and the intrinsic Zeeman effect $b$. That is, $m_1$ and $b_1$ for orbital $\tau_z=+1$ (CB) while $m_2$ and $b_2$ for orbital $\tau_z=-1$ (VB). Consequently, the four band edges at $k_z=0$ are $b_1+m_1$ (upper CB), $b_2-m_2$ (lower CB), $-(b_1-m_1)$ (upper VB), and $-(b_2+m_2)$ (lower VB), respectively.

We can then make the following substitution in the Hamiltonian
$$m\tau_z+b\sigma_z \longrightarrow  m \tau_z + \delta m + b \sigma_z+ \delta b\tau_z \sigma_z$$
where $m=(m_1+m_2)/2$, $\delta m=(m_1-m_2)/2$, $b=(b_1+b_2)/2$, and $\delta b=(b_1-b_2)/2$. Since the $\delta m$ term only shifts the Weyl cones in energy rather than reconstruction, we omit it in the following discussion. After switching the WP separation back to the $x$ direction, we arrive at the e-h asymmetric model \eqref{Ham_eh} in the main text.

To fit the {\em ab initio} results of Ref. [\onlinecite{abinitio_2}], we focus on the low-energy bands near WPs. Specifically, we perform the fitting along different $k$ directions in the range of $k_x(k_y)\in[-0.6,0.6]$ nm$^{-1}$ and $k_z\in[-0.3,0.3]$ nm$^{-1}$, with the energy $E\in[-400,400]$ meV for TaAs and TaP and $E\in[-200,200]$ meV for NbAs and NbP. We find that for relatively large $k$, it is helpful to include a parabolic band component, which is a common practice in modeling topological materials.\cite{WSM_IR_Orlita,ZrTe5_YJ_PRL,TI_Yuxuan,PRB_CXL} The parabolic band component ($\mathscr{B}$) is introduced as a second-order correction to the mass 
\begin{align*}
 M(\mathbf{k})=m+\mathscr{B}_xk_x^2+\mathscr{B}_yk_y^2+\mathscr{B}_zk_z^2,
\end{align*}
leading to an effective Hamiltonian
\begin{align}
\label{Ham_fit}
H=T(\mathbf{k})+\tau_x\hbar(\boldsymbol{\sigma}\cdot\mathbf{vk})+ M(\mathbf{k})\tau_z+b\sigma_x+\delta b\tau_z\sigma_z,
\end{align}
for carrying out the fitting.

Figure 4 shows the fitting results between the {\em ab initio} calculations of Ref. [\onlinecite{abinitio_2}] and the four-band model of \eqref{Ham_fit}. We use NbP and TaAs as examples since the other two materials in the nonmagnetic transition-metal monopnictides family have very similar band structures. As is clearly seen, our four-band model not only catches the characteristic feature of Weyl cones (lower CB and upper VB) but also describes the hybridization induced bands (upper CB and lower VB), exhibiting an excellent agreement with the {\em ab initio} results for both WP1 and WP2. It demonstrates that our four-band model can provide a realistic description of the WSMs. The extracted band parameters are listed in Table I.

\section*{Appendix II: Finite doping effects on magneto-absorption}
\begin{figure}[t!]
\includegraphics[width=8cm]{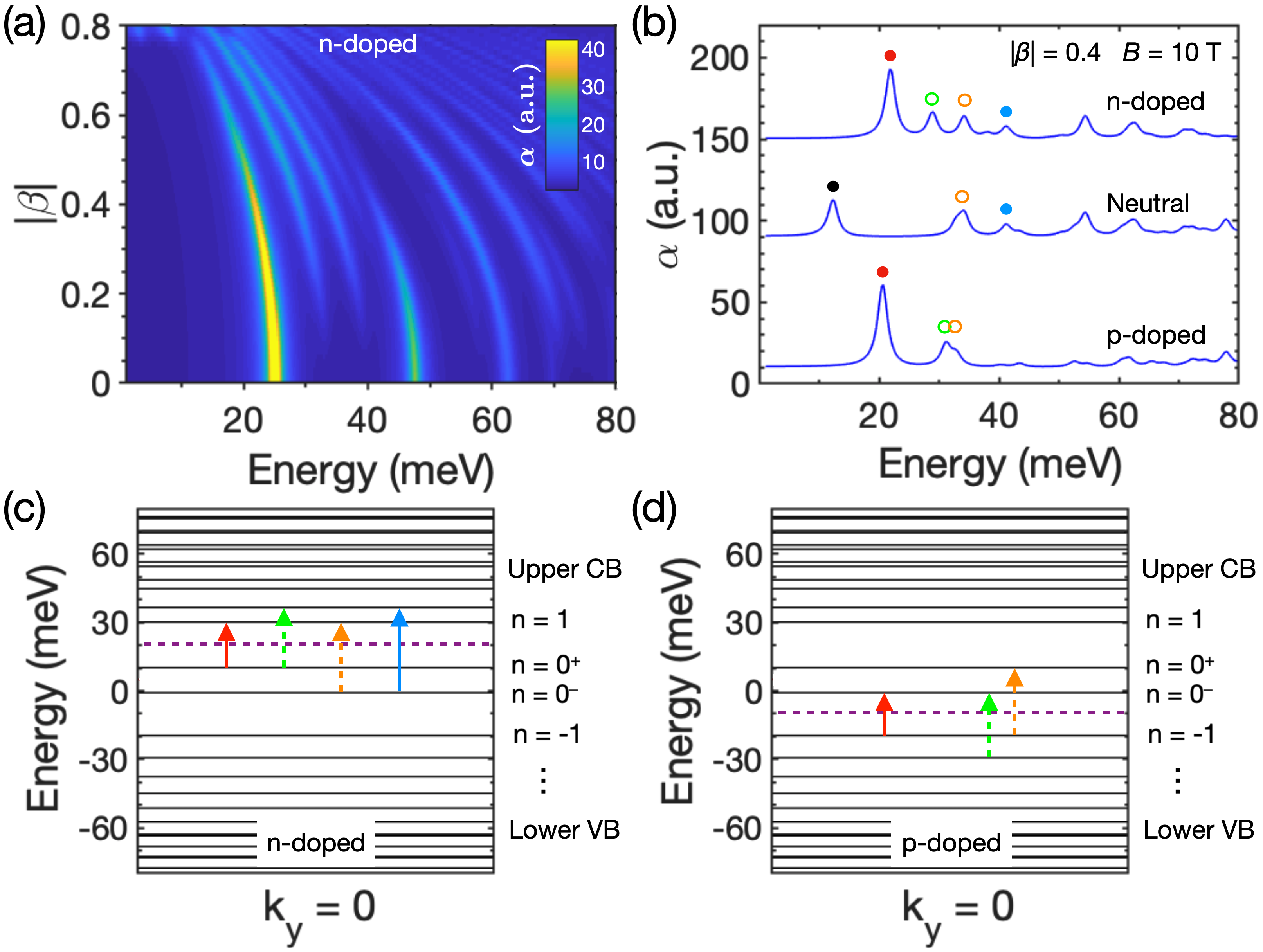}
\caption{(a) Calculated magneto-absorption spectra $\alpha(E)$ (using the tilt model) as a function of tilting angle with $E_F=20$ meV (n-doped) and $B=10$ T. (b) Comparison of the calculated magneto-absorption spectra with $|\beta|=0.4$, $B=10$ T, and $E_F=20$ meV (n-doped), 5 meV (charge neutral), and -10 meV (p-doped), respectively. (c)(d) Low-lying transitions at $k_y=0$ for the n-doped and p-doped cases in (b). The color-code corresponds to the major absorption peaks in (b). The solid (dash) arrows in (c)(d) as well as the solid (open) circles in (b) label the selection rule conserved (violated) transitions. The purple dash line in (c) and (d) indicates the Fermi level position.}
\end{figure}

In this Appendix, we discuss the finite doping effects on magneto-absorption. Figure 5(a) shows the calculated magneto-absorption spectra $\alpha(E)$ (using the tilt model) with $B=10$ T and $E_F=20$ meV (n-doped) at different tilting angles. As one can see, the spectra are similar to that in Fig. 3(a) at high energies, but the lowest transition seems significantly blueshifted and a new transition appears at slightly higher energy and $|\beta|>0.1$. These can be better seen in Fig. 5(b), where we compare the spectra at $|\beta|=0.4$ and $B=10$ T with $E_F=20$ meV (n-doped), 5 meV (charge neutral, same as Fig. 3), and -10 meV (p-doped). The doping levels are chosen in such a way to represent the quantum limit in realistic experiments. Specifically, we find that: (i) Due to the Pauli blocking effect, the $L_{0^-}\rightarrow L_{0^+}$ (black) transition in the charge neutral case disappears upon doping, replaced by a cyclotron transition (red) $L_{0^+}\rightarrow L_{1}$ ($L_{-1}\rightarrow L_{0^-}$) for the n-doped (p-doped) case. The cyclotron transition occurs at higher energy than the $L_{0^-}\rightarrow L_{0^+}$ transition, as shown in Figs. 5(c) and 5(d). (ii) As the tilt increase, a new intraband transition appears and can be identified as (green) $L_{0^+}\rightarrow L_{2}$ ($L_{-2}\rightarrow L_{0^-}$) for the n-doped (p-doped) case. These chiral transitions break the usual selection rule at zero-tilt and exhibit a similar intensity as the $L_{0^-(-1)}\rightarrow L_{1(0^+)}$ (orange) transitions. As discussed above, the presence of the $L_{0^-(-1)}\rightarrow L_{1(0^+)}$ (orange) transitions can be viewed as an indicator for the tilted Weyl bands, the $L_{0^+(-2)}\rightarrow L_{2(0^-)}$ (green) transitions can thus serve for the same purpose as well as an indicator for finite doping. Due to the e-h asymmetry, the electron-like transitions exhibit a stronger intensity than the corresponding hole-like transitions. In the p-doped case (Fig. 5(b)), the electron-like interband transitions are Pauli blocked, resulting in weaker magneto-absorption at high energies. Nevertheless, qualitatively speaking, the magneto-absorption in the n-doped and p-doped cases are similar if neglecting intensity differences and a small shift in peak energy due to e-h asymmetry.\\

\section*{data Availability}
The data that support the findings of this study are available from the corresponding author upon reasonable request.


\begin{thebibliography}{99}
\bibitem{note1}We can eliminate the $t_y$ component by rotating in the $y$-$z$ plane to align the tilt with an axis.
\end{thebibliography}
\end{document}